\documentclass{aa}
\pdfoutput=1
\usepackage{graphicx}
\usepackage{txfonts}
\usepackage{textcomp}
\usepackage{epstopdf}
\usepackage{subfigure}
\newcommand{\ds}{\ensuremath{^\circ}$\:$}
\newcommand{\dsns}{\ensuremath{^\circ}}
\newcommand{\ms}{$\textrm{m\: s}^{-1}\:$}
\newcommand{\msns}{$\textrm{m\: s}^{-1}$}
\begin{document}

   \title{Collisions of small ice particles under microgravity conditions (II): }

   \subtitle{Does the chemical composition of the ice change the collisional properties?}

   \author{C. R. Hill
          \inst{1},
          D. Hei{\ss}elmann \inst{2,3},
          J. Blum \inst{2},
          H. J. Fraser \inst{1}
          }

   \institute{The Open University, Department of Physical Sciences, Walton Hall, Milton Keynes, MK7 6AA, UK\\
              \email{catherine.hill@open.ac.uk}
         \and
             Technische Universit{\"a}t Braunschweig, Institut f{\"u}r Geophysik und extraterrestrische Physik, Mendelssohnstra{\ss}e 3, 38106 Braunschweig, Germany
         \and
            International Max-Planck Research School, Max-Planck Institute of Solar System Research, Justus-von-Liebig-Weg 3, 37077 G{\"o}ttingen, Germany\\
             }

   \date{Received 14 November 2014, Accepted 23 December 2014}


  \abstract
   {
   Understanding the collisional properties of ice is important for understanding both the early stages of planet formation and the evolution of planetary ring systems. Simple chemicals such as methanol and formic acid are known to be present in cold protostellar regions alongside the dominant water ice; they are also likely to be incorporated into planets which form in protoplanetary disks, and planetary ring systems. 
    However, the effect of the chemical composition of the ice on its collisional properties has not yet been studied.}
   {Collisions of 1.5 cm ice spheres composed of pure crystalline water ice, water with 5\% methanol, and water with 5\% formic acid were investigated to determine the effect of the ice composition on the collisional outcomes.}
   {The collisions were conducted in a dedicated experimental instrument, operated under microgravity conditions, at relative particle impact velocities between 0.01 and 0.19~\msns, temperatures between 131 and 160 K and a pressure of around $10^{-5}$~mbar.}
   {A range of coefficients of restitution were found, with no correlation between this and the chemical composition, relative impact velocity, or temperature. }
   {We conclude that the chemical composition of the ice (at the level of 95\% water ice and 5\% methanol or formic acid) does not affect the collisional properties at these temperatures and pressures due to the inability of surface wetting to take place. 
   At a level of 5\% methanol or formic acid, the structure is likely to be dominated by crystalline water ice, leading to no change in collisional properties. The surface roughness of the particles is the dominant factor in explaining the range of coefficients of restitution.}


   \maketitle
%

\section{Introduction}

Water ice is abundant in the interstellar medium and has also been observed towards dense clouds, cloud cores, protostellar regions and protoplanetary disks \citep{Oberg11}. Recent research studying deuterium-to-hydrogen enrichments in the solar system has shown that water ice was present in the solar nebula protoplanetary disk \citep{Cleeves14}. Water ice is also present in planetary ring systems such as the rings of Saturn \citep{Cuzzi78, Cuzzi80}, where its collisional properties play an important role in the structure and dynamical evolution of the rings. It is not clear whether these rings are old or young; in the former case, the rings will be residues from the formation era. The presence of water ice across star and planet forming regions may therefore influence the processes of planet formation. 

 It has been proposed that planets form from the dust in a protoplanetary disk by a process of dust aggregation \citep{weidenschilling1977,weidenschilling1980}. While this theory has gained widespread acceptance, there remains a problem in that growth between centimetre and kilometre sizes has not been demonstrated, as there is a critical velocity, reducing with particle size \citep{weidenschilling1977}, beyond which particles tend to bounce rather than stick (the so-called bouncing barrier) \citep{guettleretal2010,Zsom10, kotheetal2013}. In recent years, the presence of ice has been postulated as one possible solution to the bouncing barrier; it has been shown that ice particles have a larger adhesive and rolling friction force \citep{Gundlach11}, a higher sticking threshold \citep{Gundlach14} and reduced elasticity \citep{Hertzsch02} compared to silicate dust particles which will increase the threshold velocity for sticking. In addition, recent model simulations have shown that ice condensation could enable dust grains to grow to decimetre sizes around the snowline \citep{Ros13} and even to icy planetesimals if the initial ice grains were submicrometre-sized \citep{kataokaetal2013}.

 In the context of this paper, ring systems are also of interest. Until recently, rings had exclusively been known to exist around the four outermost solar system planets, though with the detection of two small dense rings around the Centaur (10199) Chariklo \citep{Braga-Ribas14, Duffard14}, it seems that ring systems are not exclusively present around giant planets. The rings around Saturn are the most well studied and consist predominantly of small icy bodies \citep{Cuzzi10}, between 1 cm and 10 m diameter, dominated by water ice \citep{Zebker85}. The ring dynamics are complex, influenced by perturbations by resonances between the ring particles and nearby moons, which are counteracted by frequent inelastic collisions at low relative velocities. Typical collision velocities in the unperturbed rings are well below 1 $\textrm{cm\: s}^{-1}\:$ \citep{Esposito02, Colwell09}, but gravitational perturbations may raise the average collision speed considerably. The collisions dissipate kinetic energy from the rings and thereby determine the stability of the rings \citep{Salo01, Schmidt01}; consequently, data pertaining to collisional properties and outcomes are vital to model efforts which are used to understand ring instabilities, wakes and overall structure, including the very strongly confined heights observed in Saturnian and other ring systems \citep{Colwell06, Colwell07, Hedman07, Thomson07}.
 
 While water is abundant in planet forming regions, many other molecules have also been detected in protostellar regions. Consequently, the icy mantles on dust grains in protoplanetary disks are likely to be composed of water dominated ices which may also contain a plethora of "contaminants", including many simple molecules such as CO, CO$_2$, CH$_3$OH, CH$_4$, NH$_3$ and HCOOH (e.g. \cite{Pontoppidan04}, \cite{Aikawa12}, \cite{Oberg11_2}, \cite{Noble13}). Recent work has suggested that the most abundant volatile species in planets of solar composition are H$_2$O, CO, CO$_2$, CH$_3$OH and NH$_3$ \citep{Marboeuf14}. It is also possible that some of these species may be present in planetary rings such as those of Saturn. While Saturn's rings are composed of mainly water ice, there is evidence that other species are present \citep{Poulet03, Cuzzi09} and it is not unreasonable to suppose that some of those may include the above molecules. This provides the main motivations for this paper: firstly to study the effects of minor chemical pollutants on collisions between icy particles at low velocities, and secondly to understand how the outcomes of these collisional processes affect processes of planet formation and ring system evolution.

The work on this paper builds on our previous publication (\cite{Hill14}, henceforth Paper I) in which the collisions of millimetre-sized ice particles (both spheres and irregularly shaped fragments, diameters 4.7-10.8~mm) were studied under microgravity conditions on a parabolic flight. The main parameters studied were the coefficient of restitution, $\varepsilon$, and the impact parameter, \textit{b}. The coefficient of restitution is the ratio of the relative velocities of the particles after and before the collision, which is related to the translational kinetic energy lost in the collision:  \begin{equation} \varepsilon = \frac{v_{\textrm{a}}}{v_{\textrm{b}}} \end{equation} where a and b denote after and before the collision respectively. The impact parameter is the distance of closest approach of the two particles perpendicular to their relative velocity vector. Henceforth we use the normalised impact parameter \textit{b/R}, which is the impact parameter, \textit{b}, normalised to \textit{R}, the distance between the centre of masses of the two particles at the point of collision, i.e. the sum of the two radii for spherical particles.

The relative collision velocities ranged from 0.26 to 0.51~\msns. The coefficients of restitution were spread evenly between 0.08 and 0.65 with no dependence of this property on either impact velocity or normalised impact parameter. The spread of coefficients of restitution was attributed to the surface roughness of the particles. This corroborates previous work by \cite{Heisselmann10}, where collisions of 1.5~cm (diameter) ice spheres were studied under microgravity conditions. In this case, the impact velocities were between 0.06 and 0.22~\ms and the coefficient of restitution was spread evenly between 0.06 and 0.84. The difference between the ranges of coefficients of restitution was thought to be due to the difference in velocity ranges between the two studies.

 The range of coefficients of restitution and lack of dependence on impact velocity and normalised impact parameter is in contrast to previous studies of ice collisions which show a decrease in coefficient of restitution with increasing impact velocity (\cite{Bridges84}, \cite{Hatzes88}) and very little loss of kinetic energy for glancing collisions (i.e. \textit{b/R} $\sim$ 1) \citep{Supulver95}. The reason for this was thought to be the difference in surface roughness; while the ice surfaces in the work of \cite{Bridges84}, \cite{Hatzes88} and \cite{Supulver95} were smooth, the ice surfaces in the studies of \cite{Heisselmann10} and Paper I were rough and anisotropic. This explains the ranges of coefficient of restitution and the lack of dependence on impact velocity and normalised impact parameter and is in accordance with the work of \cite{Hatzes88} where both frost and roughened surfaces were found to reduce the coefficient of restitution. For a full comparison of the experimental conditions in these studies, we refer the reader to Paper I.

An inherent limitation in the studies of ice collisions to date (as related to planet formation and planetary ring systems) is their focus on pure water ice. It is possible that the presence of other chemicals might change the collisional properties; \cite{Bridges96} studied the sticking forces between frost coated water ice at atmospheric pressure and temperatures between 110-150~K and discovered that methanol frosts had stronger sticking forces than water frosts. The melting point of methanol is 176~K at atmospheric pressure so it is possible that the increased sticking forces are due to surface melting. This is the same mechanism that leads to ice aggregation in atmospheric clouds \citep{Hobbs65}, leading to increased aggregation around the melting point of water \citep{Hobbs74}. In protoplanetary disks and planetary rings, the temperatures and pressures are too low for this process to dominate. Both methanol and formic acid have been detected in protostellar regions in the solid phase (\cite{Pontoppidan04}, \cite{Oberg11_2}, \cite{Aikawa12}, \cite{Schutte99}, \cite{Zasowski09}) at the 1-30\% abundance level relative to water ice for methanol and 1-5\% level for formic acid \citep{Boogert08}. We have therefore chosen to focus this study on icy particles  formed by freezing either pure water or 5\% methanol or formic acid in water, to investigate the effects of the chemical composition of ice on its collisional properties. While the eventual aim is to investigate the effects of the chemical composition on planet formation, the form of ice used here (solid ice spheres) is more relevant to planetary ring systems than protoplanetary disks, where the ice is likely to be in the form of porous agglomerates (e.g. \cite{kataokaetal2013}). Therefore the results will be more immediately applicable to planetary rings, but it is hoped that they will be a starting point for discussion of ice in protoplanetary disks as well.


\section{Experimental details}

\begin{figure}
   \centering
   \includegraphics[width=\hsize]{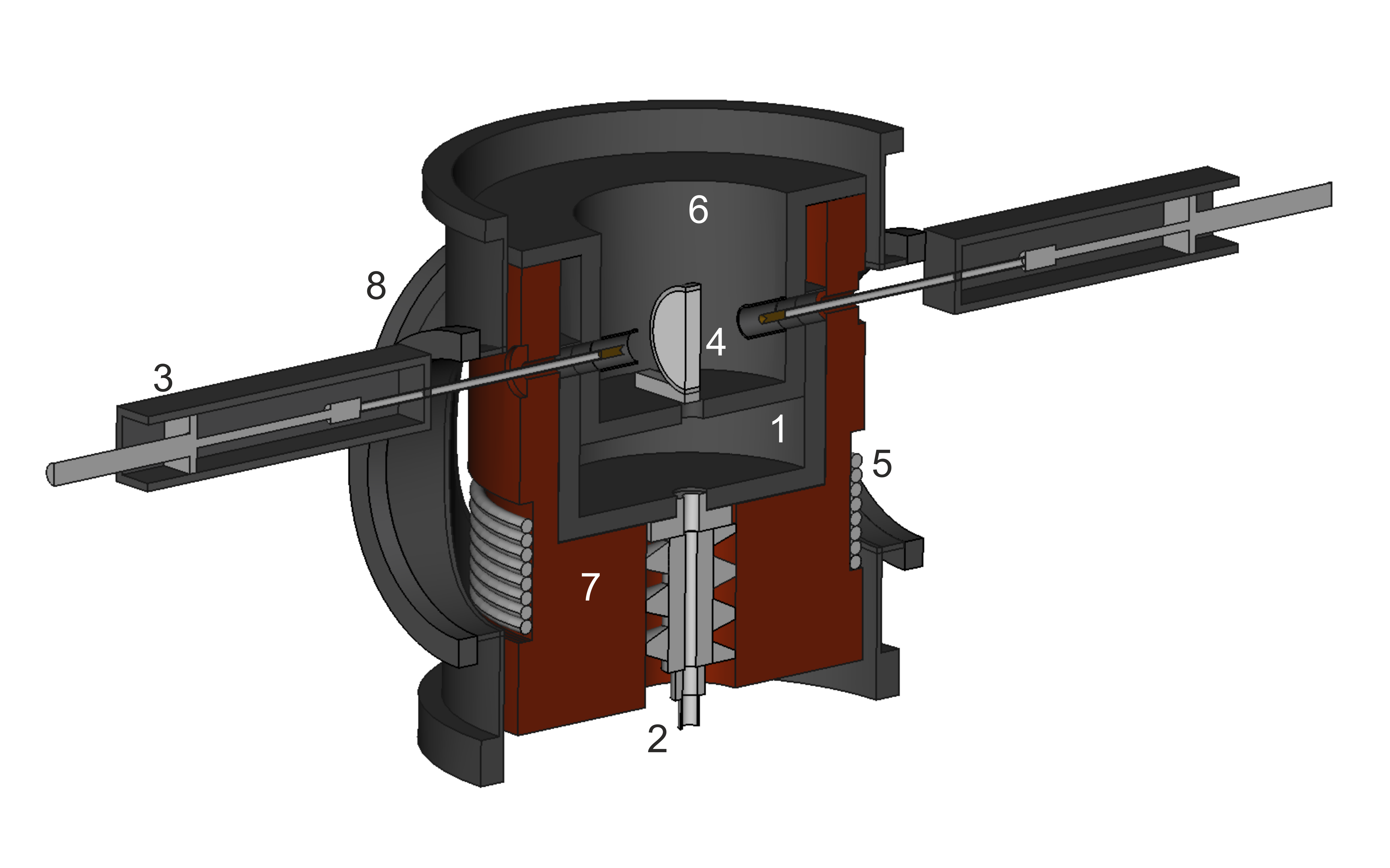}
      \caption{Computer aided design schematic of the experimental setup (adapted from \cite{Salter09}. The particles are stored in an aluminium colosseum (1) which is rotated (2) to line up two colosseum holes with two diametrically opposed motorised pistons (3) which collides the particles either with a 52~mm diameter ice target (4) placed at the centre of the collision volume or with each other when the target is removed. The system is passively cooled by passing liquid nitrogen through a  cooling ring (5) prior to take off. The particles are kept within the colosseum and protected from radiative heating by an aluminium shield (6), and a 45~kg copper block (7) acts as a thermal reservoir, keeping the system cool during the flight. The entire set up is enclosed within a vacuum chamber (8).
              }
         \label{FigExp}
   \end{figure}

The experimental setup has previously been discussed in detail (\cite{Salter09} and Paper I) and is summarised again here for completeness (see Fig.~\ref{FigExp}). The experiment was designed and built to carry out low velocity particle collisions on parabolic flights. Prior to take off, the particles are loaded into a pre-cooled (to around 77~K) rotating double helix particle reservoir, or colosseum (1). This is rotated (2) during the flight to line up two diametrically opposed ice particles with motorised pistons (3) which accelerate the particles to a pre-defined constant velocity directed towards the centre of the collision volume, where they collide in perfect free fall conditions either with a target (4) or with each other. The system is cryogenically cooled to temperatures around 77~K by passing liquid nitrogen through a cooling ring (5)  prior to take off and is kept cool during flight with the use of a U-shaped aluminium heat shield (6) and a 45~kg copper block (7) on which the particle reservoir sits. The copper acts as both a thermal reservoir and a cryo-pump for any residual gas. The entire set up is encased inside a vacuum chamber (8), with a typical residual gas pressure of $10^{-5}$~mbar. The pistons enter the collision volume through two differentially pumped feedthroughs. A collision is initiated by accelerating the pistons (which are controlled by Labview software) to a maximum velocity of between 0.01 and 0.1~\ms. The pistons make contact with the particles, are brought to a sharp stop a few millimetres later, and then are immediately retracted to their starting positions. The piston tips remain within the cooled region at rest, meaning that they are at the same temperature as the particles and do not induce particle heating or surface melting during collision. A high speed camera with a frame rate of 107 frames per second was combined with prism optics to capture two views of the collision 48.8\ds  apart.

\begin{figure}[h]
   \centering
   \includegraphics[width=\hsize]{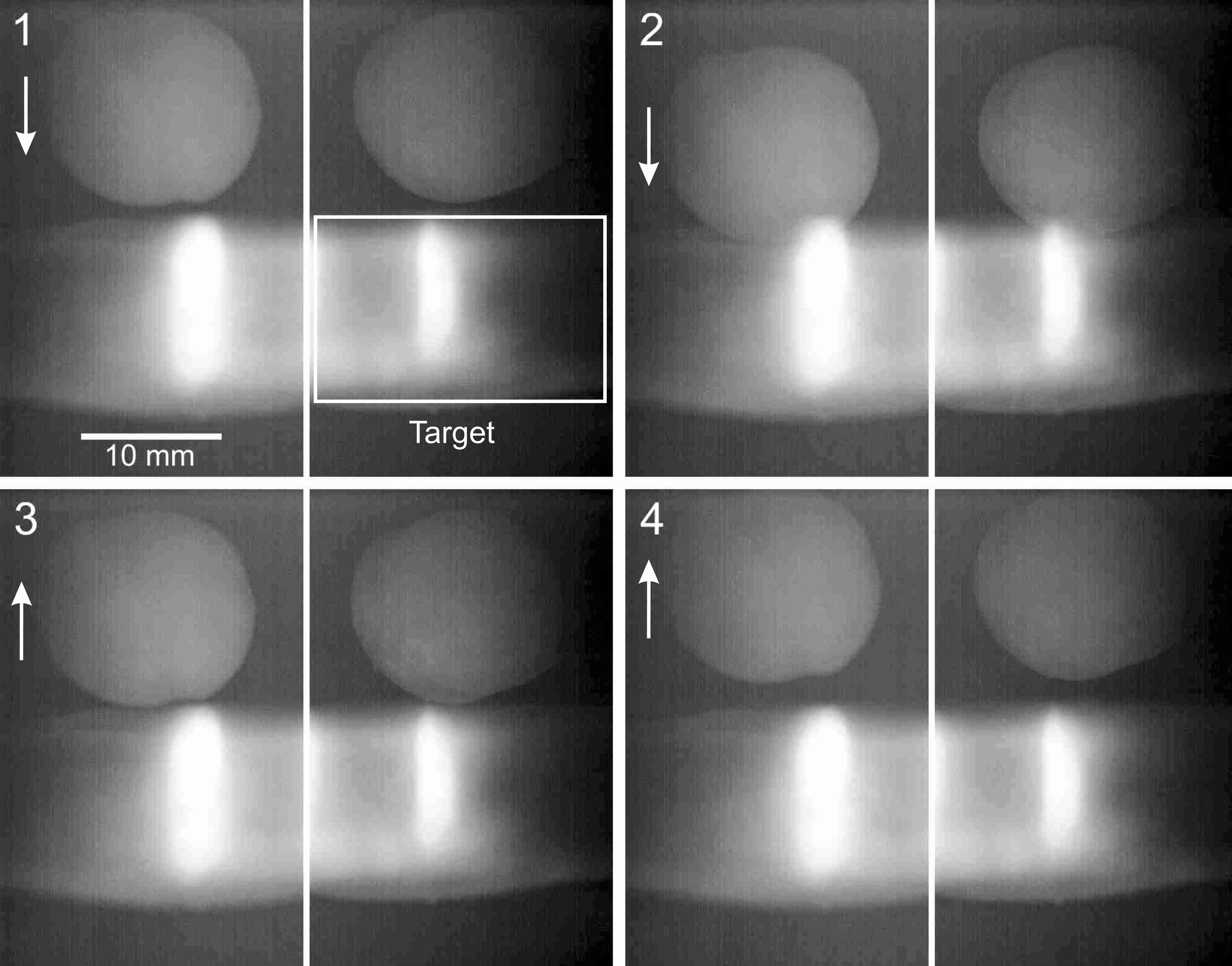}
      \caption{Image sequence of an ice sphere containing 5\% formic acid colliding with a pure ice target (indicated) at 90\ds  to the direction of travel at a velocity of 0.08 \msns. The images were captured using beam splitter optics with the view on the left separated from the view shown on the right by 48.8\dsns. The images are taken 4/107 s apart.
              }
         \label{FigImSeq}
   \end{figure}

The particles used in this experiment were ice spheres of 1.5 cm diameter. Some were composed of pure, distilled water, and others were composed of a 5\% solution of methanol or formic acid in distilled water. The particles were produced using spherical moulds and a standard kitchen freezer. Making ice particles in this way produces crystalline hexagonal ice with a rough, anisotropic surface \citep{Heisselmann10}. The particles were removed from the freezer and placed directly into a container of liquid nitrogen situated within the chamber. Once the particles had reached 77 K, they were individually loaded into the pre-cooled sample reservoir. Surface frosting was minimised by pre-cooling the particle reservoir to $\sim$77 K prior to loading and evacuating the chamber to a base pressure of around $10^{-5}$~mbar immediately after loading while the ice particles were still outgassing N$_2$, thereby preventing frost from forming on the particle surfaces. The target was composed of pure crystalline water ice and was placed in the centre of the chamber immediately prior to evacuation. The target was placed in a different orientation with respect to the chamber for each flight, giving access to different normalised impact parameters. The normalised impact parameter is related to the angle of the target to the direction of travel ($\theta$) by the following equation: \begin{equation} b/R = cos(\theta) \end{equation} On the first flight of this campaign, the target was at 90\ds to the direction of travel of the particles, giving \textit{b/R} = 0 - completely head on collisions. The second flight had the target at  60\ds  to the direction of travel, giving \textit{b/R} = 0.5. The third flight had the target mounted at 30\ds  to the direction of travel, giving $b/R = \sqrt{3}/2$. Technical problems prevented collection of data in this configuration and so particle-particle collisions were performed instead. It was therefore necessary to open the chamber during the flight to remove the target. The particles remained within the sample reservoir and hence their exposure to the atmosphere of the plane was minimal. It is unlikely that significant frost formation could occur on the particle surfaces under these conditions and comparison with previous particle-particle collisions conducted under similar conditions (but without the opening of the chamber) shows that this assumption is correct (see Section~\ref{Results_COR} for a comparison of our data with that of \cite{Heisselmann10}). Fig.~\ref{FigImSeq} shows an example collision of an ice sphere containing 5\% formic acid with the ice target at 90\ds  to the direction of travel.

Table~\ref{collisiontable2} shows a break down of all the collisions successfully performed during this campaign. The numbers of collisions suitable for analysis and unsuitable for analysis are given. Residual acceleration of the plane (when the quality of microgravity is poor and the camera moves horizontally with respect to the particles as a result) was a particular problem on the second day of the campaign (target at 60\ds) and hence only 8 collisions suitable for analysis were obtained. Poor quality hampered image analysis efforts in a few cases and there was one case of fragmentation. This is thought to be related to the way in which the ice spheres are formed rather than a real collisional outcome.
Table~\ref{collisiontable1} shows a break down of the collisions suitable for analysis, the results of which are presented in Section~\ref{Results_COR}.

\begin{table}[h]
\caption{A break down of the collisional outcomes observed in this study. The angles refer to the angle between the target and the direction of travel of the particles. The collisions that were suitable for analysis are detailed in Section~\ref{Results_COR}. The other collisions were excluded from the analysis due to residual acceleration (poor quality of microgravity), poor image quality and fragmentation. }             
\label{collisiontable2}      
\centering                          
\begin{tabular}{c c c}        
\hline\hline                 
 & Collisional outcome & Number of collisions  \\    
\hline                        
   Target at 90\ds & Suitable for analysis & 35   \\      
    & Residual acceleration & 1 \\
     & Poor quality & 4 \\
     & Fragmentation & 1 \\
   \hline
   Target at 60\ds & Suitable for analysis & 8   \\      
    & Residual acceleration & 41 \\
     & Poor quality & 5 \\
     & Fragmentation & 0 \\
   \hline
   No target & Suitable for analysis & 15   \\      
    & Residual acceleration & 3 \\
     & Poor quality & 7 \\
     & Fragmentation & 0 \\
     \hline
    & Total & 120 \\

\hline                                   
\end{tabular}
\end{table}

\begin{table}[h]
\caption{A break down of analysed collisions in terms of chemical composition. The angles refer to the angle between the target and the direction of travel of the particles. The ice spheres were composed of pure hexagonal crystalline water ice, 5\% methanol in water and 5\% formic acid in water.}             
\label{collisiontable1}      
\centering                          
\begin{tabular}{c c c c c}        
\hline\hline                 
Collision type & Pure water & 5\%  & 5\%  & Total \\    
 &  &  methanol &  formic acid &  \\
\hline                        
   Target at 90\ds & 7 & 10 &18 & 35  \\      
   Target at 60\ds & 8 & 0& 0 & 8  \\
   No target & 2 & 8 & 5 & 15\\
   \hline
    & & & Total & 58 \\

\hline                                   
\end{tabular}
\end{table}

\section{Analysis methodology}

The data analysis was performed as follows. The particles were manually tracked frame by frame in both views of the collision; the co-ordinates from the two views were combined using a transformation algorithm to give orthogonal (\textit{x}, \textit{y}, \textit{z}) co-ordinates. Linear fits in each dimension yielded particle trajectories from which velocities before and after the collision and hence coefficients of restitution were calculated. The velocities were resolved into components normal and tangential to the colliding surfaces to give normal and tangential coefficients of restitution. For the particle-particle collisions, normalised impact parameters (\textit{b/R}) were also calculated from particle trajectories. Where errors are shown, they were calculated using standard methods for the propagation of errors.

\section{Results and discussion}

\subsection{Coefficient of restitution}
\label{Results_COR}
\begin{figure}[h]
   \centering
   \includegraphics[width=\hsize]{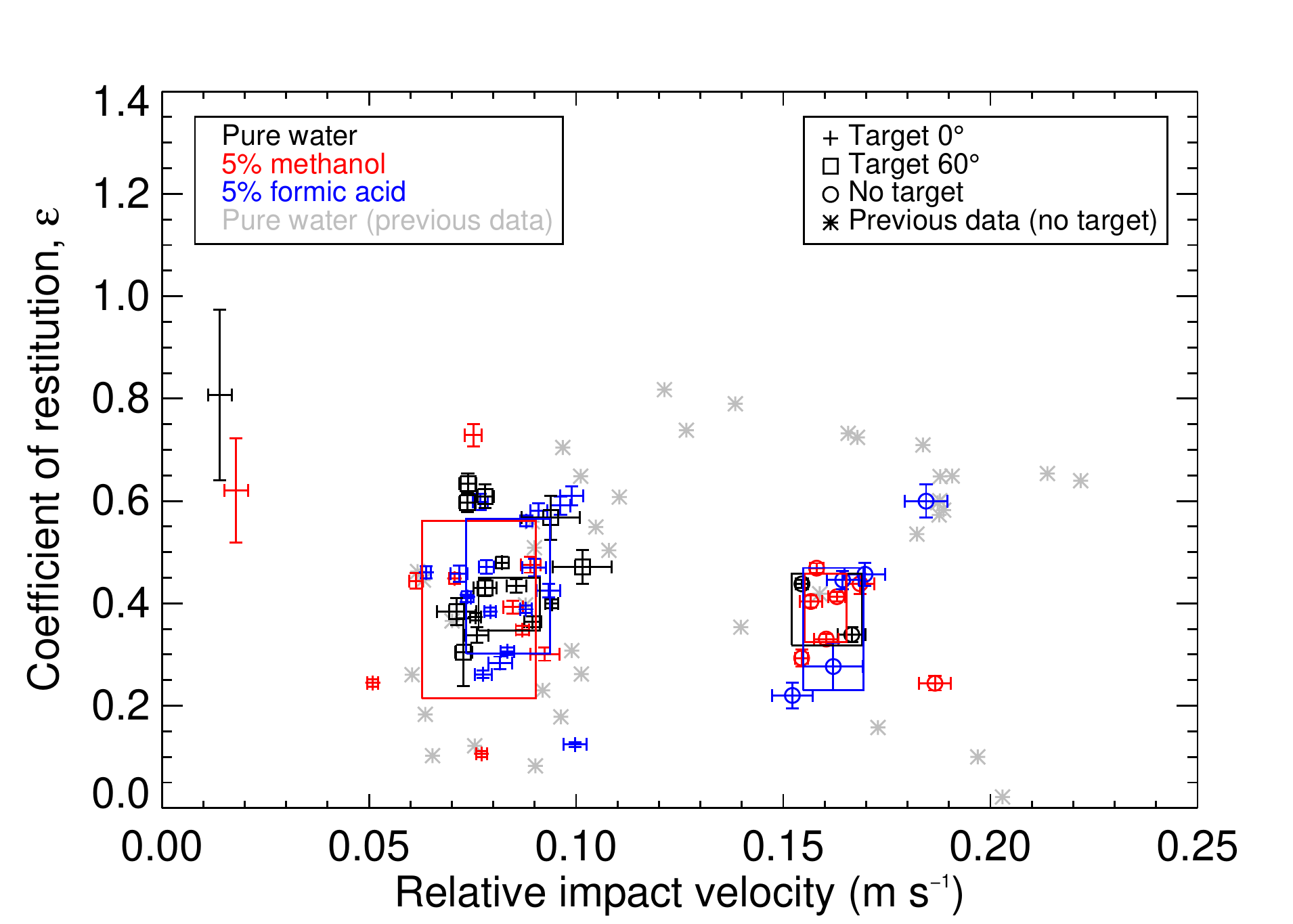}
      \caption{Coefficient of restitution as a function of impact velocity for collisions of 1.5~cm (diameter) ice spheres with a target at 90\ds to the direction of travel (\textit{b/R} = 0, shown by crosses), a target at 60\ds to the direction of travel (\textit{b/R} = 0.5, shown by squares) and no target (both this data (shown by circles) and data previously reported by \cite{Heisselmann10} (shown by stars)). The chemical composition of the spheres are indicated with colours (black for water, red for methanol, blue for formic acid and grey for the previous data). The rectangles (black for pure water, red for 5\% methanol in water and blue for 5\% formic acid in water) show the data encompassed by one standard deviation of the mean.
              }
         \label{FigAll}
   \end{figure}

Bouncing was the only collisional outcome observed in this study, corroborating previous results in Paper I and also those of \cite{Heisselmann10}. If ice particles have a critical velocity for the onset of bouncing, it will be below 0.01 \ms (the lowest velocity collision in the current study) for particles of this size. Fig.~\ref{FigAll} shows the coefficients of restitution ($\varepsilon$) as a function of relative impact velocity for this study compared the previous results of \cite{Heisselmann10} which were for collisions of pure water ice spheres of identical size. Apart from two outliers around 0.01~\msns, the data in the current study lies virtually within the range of the previous data. The two distinct velocity groupings in our data are because of the different relative velocities obtained by colliding a particle with a stationary target (the lower velocity region) and colliding a particle with another moving particle (the higher velocity region). The velocity range is similar for both studies (0.05-0.19~\ms for the current study (disregarding the two outliers around 0.01~\msns)) and 0.06-0.22~\ms for the previous study). The values of $\varepsilon$ are spread between 0.08 and 0.81 for our data and between 0.02 and 0.84 for the previous data; the range is virtually identical for the current and previous results. There is no apparent difference between the collisional properties of the different chemical compositions. The similarity of the target data with the previous particle-particle collision data shows that colliding a sphere with a larger body has the same effect as colliding it with another sphere, a result which is not surprising considering the similarities in contact area during collision in both cases. Finally, the similarity of these particle-particle collisions with the previous work reiterates the fact that the brief opening of the chamber during the flight to remove the target did not affect the results; if significant frosting of the particle surfaces had occurred during this time, the coefficients of restitution would be lower \citep{Hatzes88}.

 Where there was sufficient data, rectangles have been plotted to show how many points lay within one standard deviation of the mean (ignoring any outliers). These rectangles have been plotted independently for each experimental setup. For the target at 90\ds and for particle-particle collisions, there is a good overlap between the rectangles, demonstrating that there is no statistical difference between the datasets. We therefore conclude that the presence of methanol and formic acid in the ice at a level of 5\% does not change the coefficient of restitution. For the target at 60\dsns, technical difficulties prevented the capture of useful collisions for any other material than pure water, meaning that the effect of chemical composition cannot be tested for this flight.

At this point it is instructive to consider why the inclusion of methanol and formic acid does not noticeably affect the collisional properties of the ice spheres. As mentioned in the introduction, methanol frosts are much stickier than water frosts between 110 and 150 K at atmospheric pressure \citep{Bridges96}; this is most likely because of surface wetting near the melting point. However, surface wetting is not possible at the low pressures within our chamber for either methanol or formic acid. It is also useful to consider where the methanol is located within our ice spheres. Considering the methanol-water phase diagram (\cite{Takaizumi97}, \cite{Deschamps10}), at 255 K (the temperature of a standard kitchen freezer), the water will have frozen but the methanol will remain as a liquid. As the freezing process will freeze from the outside in, it is possible that all of the methanol is at the centre of the ice sphere due to exclusion from the water ice as it freezes, meaning that the surface is essentially pure water ice which would explain the similarity in collisional properties. Formic acid freezes at 282 K, which is higher than the freezing point of water, although it is likely that the presence of the water will depress the freezing point of formic acid. At a concentration of 5\%, we speculate that the formic acid will be distributed throughout the ice sphere due to the similarity in freezing points. Regardless of where in the spheres the methanol and formic acid is present, at a level of 5\%, the overall structure is likely to be dominated by crystalline water ice. It is likely for this reason that the collisional properties show no dependence on the composition of the ice.

To determine the strength of any correlation between the variables, the linear Pearson correlation coefficient was computed: \begin{equation}r=\frac{\sum_{i=0}^n (y_i-\bar{y})(x_i-\bar{x})}{\sqrt{\sum_{i=0}^n (y_i-\bar{y})^2}\sqrt{\sum_{i=0}^n (x_i-\bar{x})^2}} \end{equation} The number of points is given by \textit{n} and the mean values of \textit{x} and \textit{y} are given by $\bar{x}$ and $\bar{y}$ respectively. This gives a measure of the strength of correlation between two variables, x and y, with a value of 1/-1 indicating perfect positive/negative correlation and a value of 0 indicating no correlation. The value returned is only significant where there is a significant number of data points and so only cases that meet this condition will be discussed. For the target at 90\dsns, the data for pure water, methanol and formic acid shows no correlation between coefficient of restitution and impact velocity (values of 0.21, 0.06 and 0.05 respectively). The same is true for pure water with the target at 60\ds (correlation coefficient of 0.13) and for methanol with no target (correlation coefficient of 0.47); the rest of the datasets contain an insufficient number of points for any correlation to be statistically significant. These values indicate that the coefficient of restitution does not depend significantly on impact velocity which corroborates our previous work in Paper I and the work of \cite{Heisselmann10}. Here, as before, the surface roughness is likely to be the most important factor in explaining the range of coefficients of restitution.

\begin{figure*}
\centering
\includegraphics[width=0.49\textwidth]{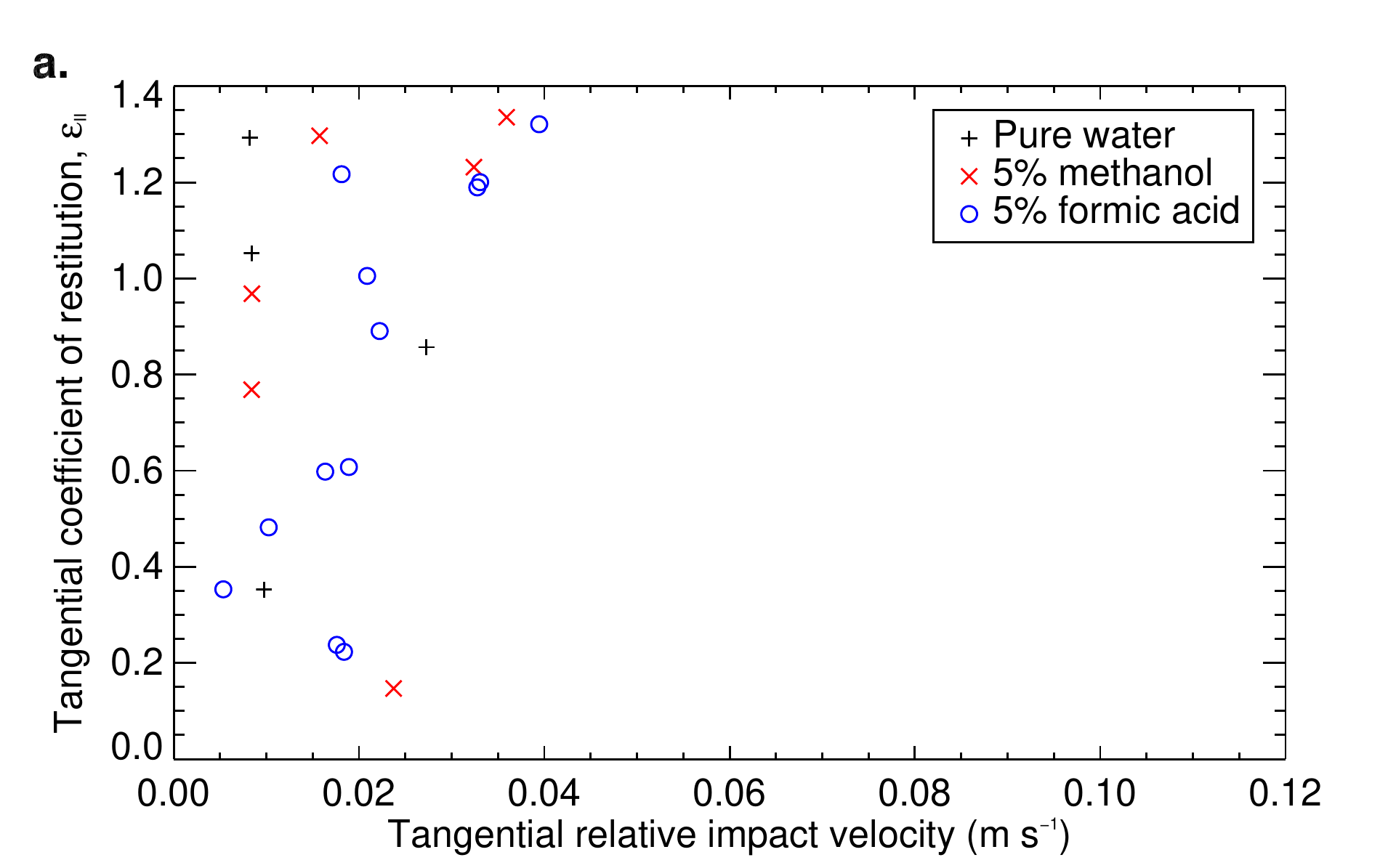}
\includegraphics[width=0.49\textwidth]{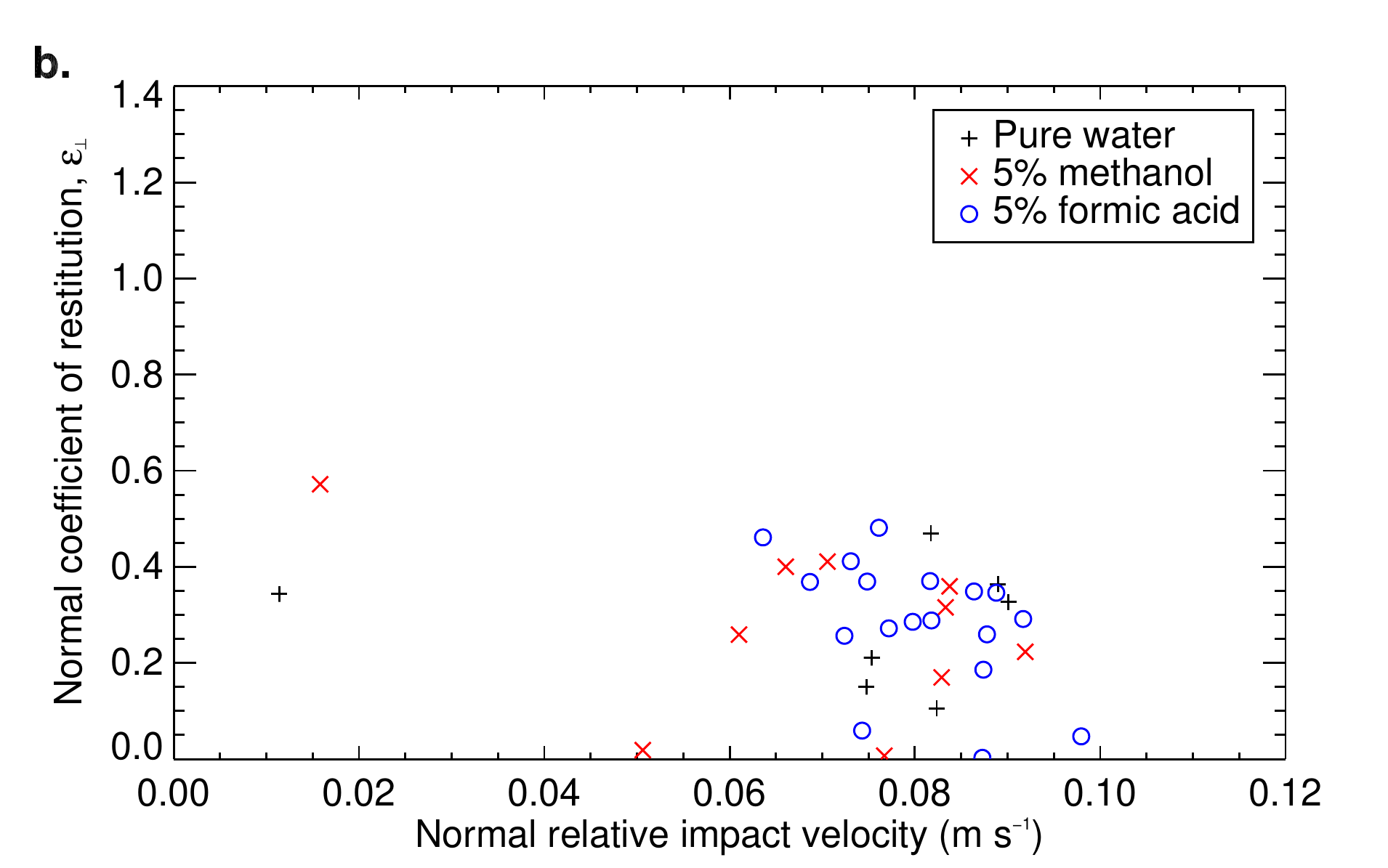}
\includegraphics[width=0.49\textwidth]{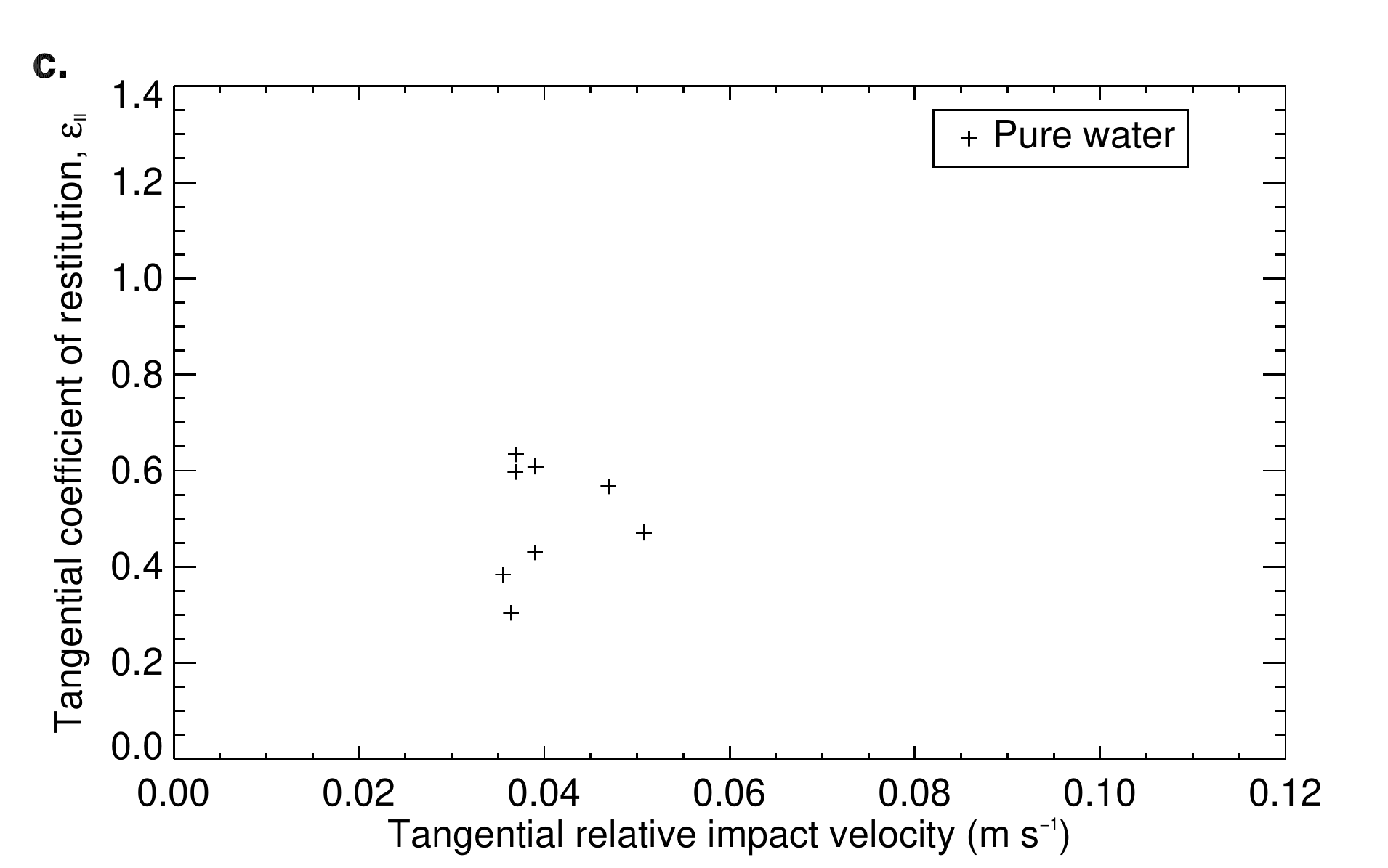}
\includegraphics[width=0.49\textwidth]{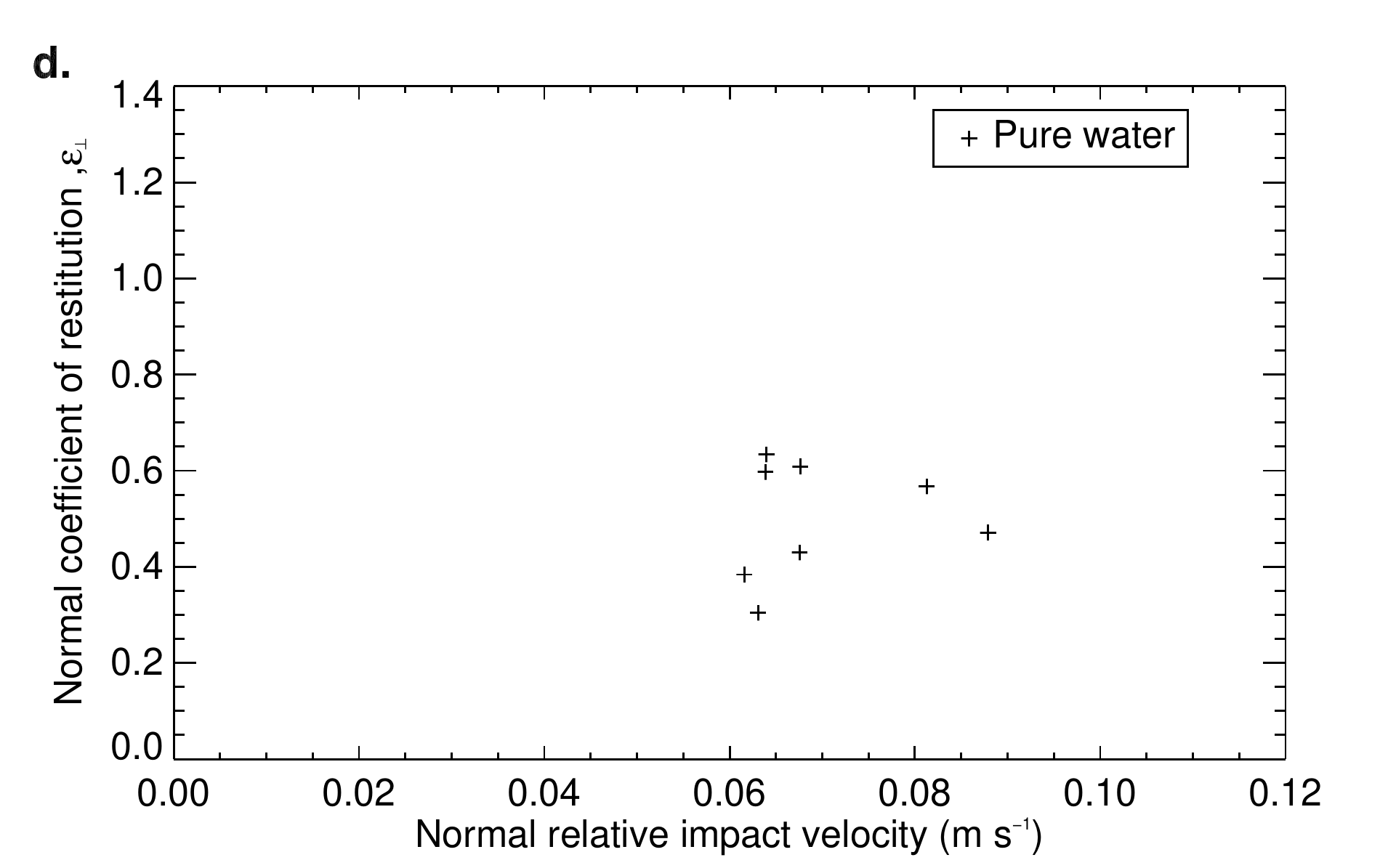}
\includegraphics[width=0.49\textwidth]{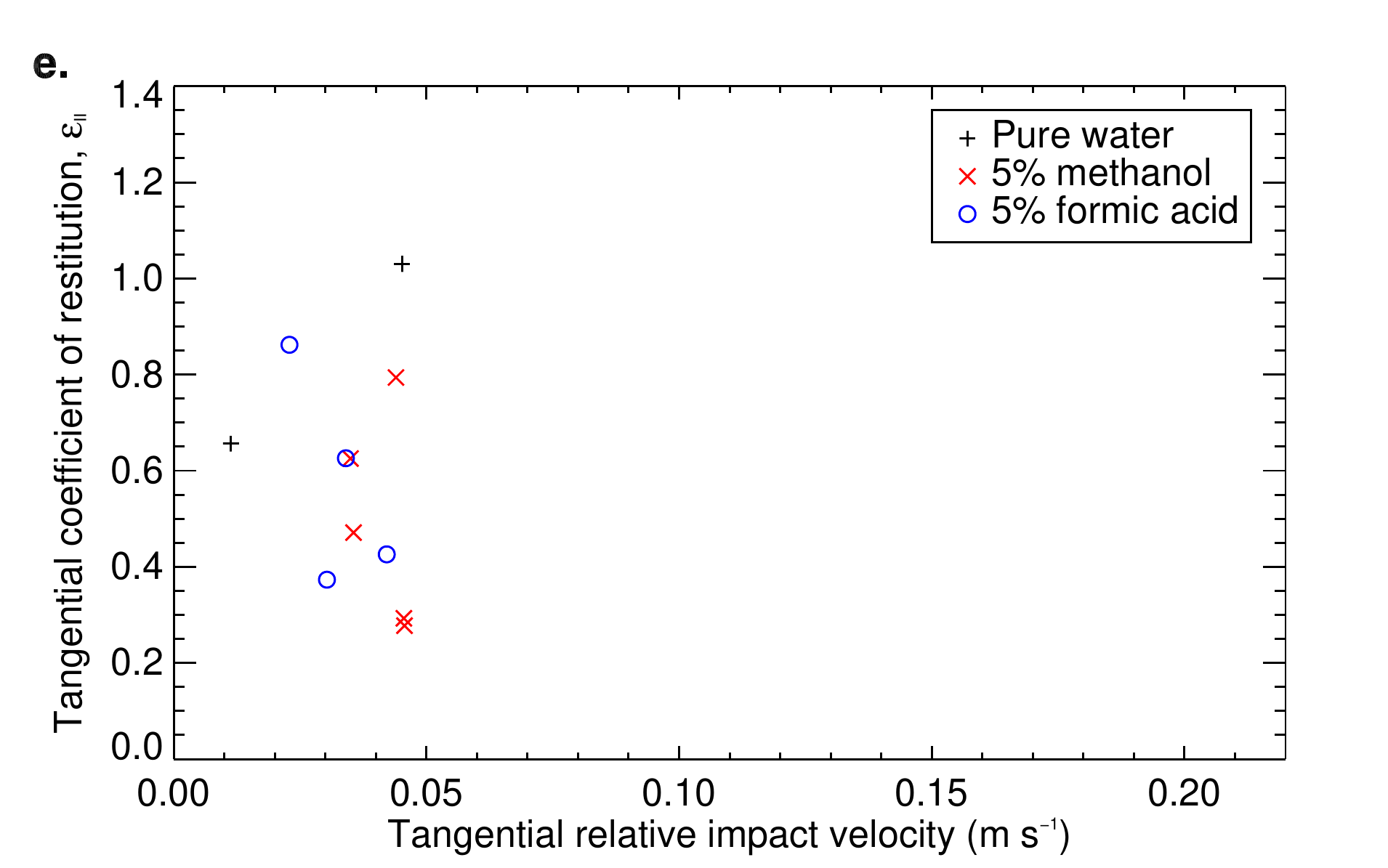}
\includegraphics[width=0.49\textwidth]{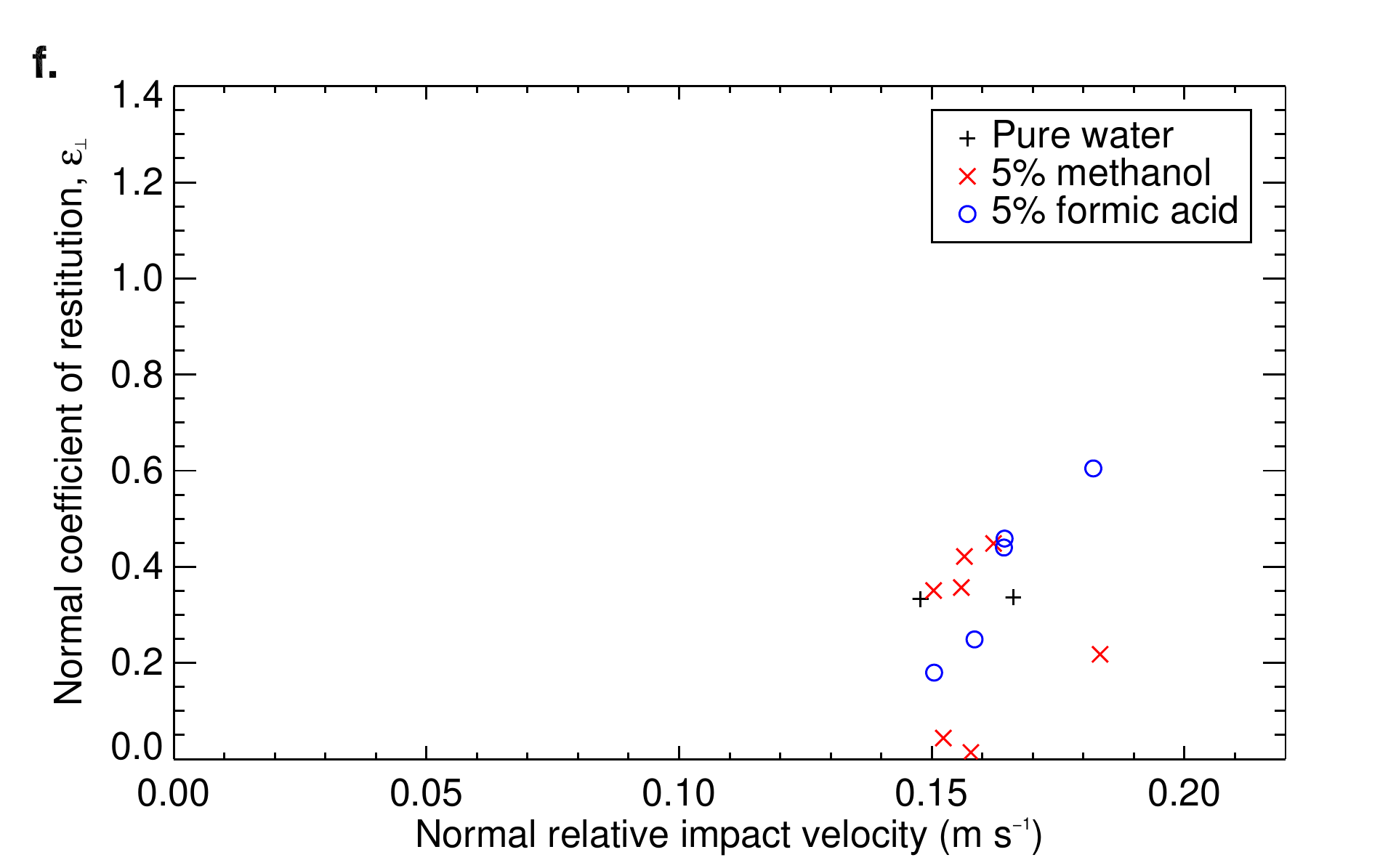}
\caption{Coefficients of restitution normal and tangential to the colliding surfaces as a function of the normal and tangential impact velocity respectively. The chemical compositions are indicated by different colours and symbols: black plus symbols for pure water ice, red crosses for 5\% methanol in water, and blue circles for 5\% formic acid in water. \textbf{a.} Tangential coefficient of restitution for the target at 90\ds to the direction of travel (\textit{b/R} = 0). \textbf{b.} Normal coefficient of restitution for the target at 90\ds to the direction of travel (\textit{b/R} = 0). \textbf{c.} Tangential coefficient of restitution for the target at 60\ds to the direction of travel (\textit{b/R} = 0.5). \textbf{d.} Normal coefficient of restitution for the target at 60\ds to the direction of travel (\textit{b/R} = 0.5). \textbf{e.} Tangential coefficient of restitution for particle-particle collisions (\textit{b/R} = 0.00-0.33). \textbf{f.} Normal coefficient of restitution for particle-particle collisions (\textit{b/R} = 0.00-0.33).}
\label{Fig_CORpape}
\end{figure*}

Fig.~\ref{Fig_CORpape} shows the coefficients of restitution tangential and normal to the colliding surfaces as a function of tangential and normal impact velocity respectively. Calculating these gives further information about the distribution of energy after the collision. The normal coefficient of restitution gives information about the rebound of the particles and the tangential coefficient of restitution gives information about particle scattering. The data for the target at 90\ds (\textit{b/R}=0) and the data for particle-particle collisions (\textit{b/R}=0.00-0.33) both have a small component of tangential impact velocity compared to the normal component (Fig.~\ref{Fig_CORpape}a, b, e, f), which is to be expected for head on (or nearly head on) collisions. The tangential coefficients of restitution are spread from around 0.1 to 1.3 and from around 0.2 to 1.1 for particle-particle collisions. The spread for the normal coefficients of restitution is much less, from around 0 to 0.6 for the target at 90\dsns, and from around 0 to 0.7 for particle-particle collisions. This shows that head on collisions can result in both scattering and rebound, with more energy going into scattering than rebound on average. For the target at 60\ds (\textit{b/R}=0.5), there is a much larger tangential component to the impact velocity than for the target at 90\dsns, although the normal component to the impact velocity is still larger (Fig.~\ref{Fig_CORpape}c and d). The spreads of tangential and normal coefficients of restitution are very similar in this case, both around 0.3-0.7. This demonstrates that for \textit{b/R}=0.5, the distribution of energy into rebound and scattering is roughly equal. In all cases, the chemical composition of the ice does not affect the distribution of translational kinetic energy.  

Considering the highest value of $\varepsilon$ obtained in this study, a minimum of 29\% of the particles' translational kinetic energy ($1-\varepsilon^2$ with $\varepsilon = 0.84$) is lost in the collision.  From the image sequences of the collisions, it is clear that some of the energy is converted into rotational energy. A break down of the particle rotation is shown in Table~\ref{Rotationtable}. The overall values are given but dividing the spheres by chemical composition does not significantly alter the proportions. Due to the spherical shape of the particles (and hence the lack of distinguishing marks), it was not possible to extract more quantitative rotational information.

\begin{table}[h]
\caption{The percentage of particles that rotate and do not rotate before and after the collision.}             
\label{Rotationtable}      
\centering                          
\begin{tabular}{c c c c}        
\hline\hline                 
      & Rotates (\%) & Does not rotate (\%) & Unclear (\%) \\    
\hline                        
   Before & 4 &	89 &	7\\
   After & 71 &	10 &	19\\
\hline                                   
\end{tabular}
\end{table}

The majority of the particles were fired from the pistons without any observable rotation (89\%). The majority of the particles did rotate after the collision (71\%) but 10\% did not. It is clear from this and previous results in Paper I that rotation cannot account for all of the particles' energy loss. A computational study by \cite{Zamankhan10} showed that most translational kinetic energy in icy grain collisions is dissipated due to surface fracturing. We cannot confirm or refute this here because such surface fracturing would not be visible in our images. It is also possible that energy is converted into heat leading to desorption of surface material, which again is not detectable in our experiment and is beyond the scope of this paper.

\subsection{Effect of temperature}

\begin{figure}[h]
   \centering
   \includegraphics[width=\hsize]{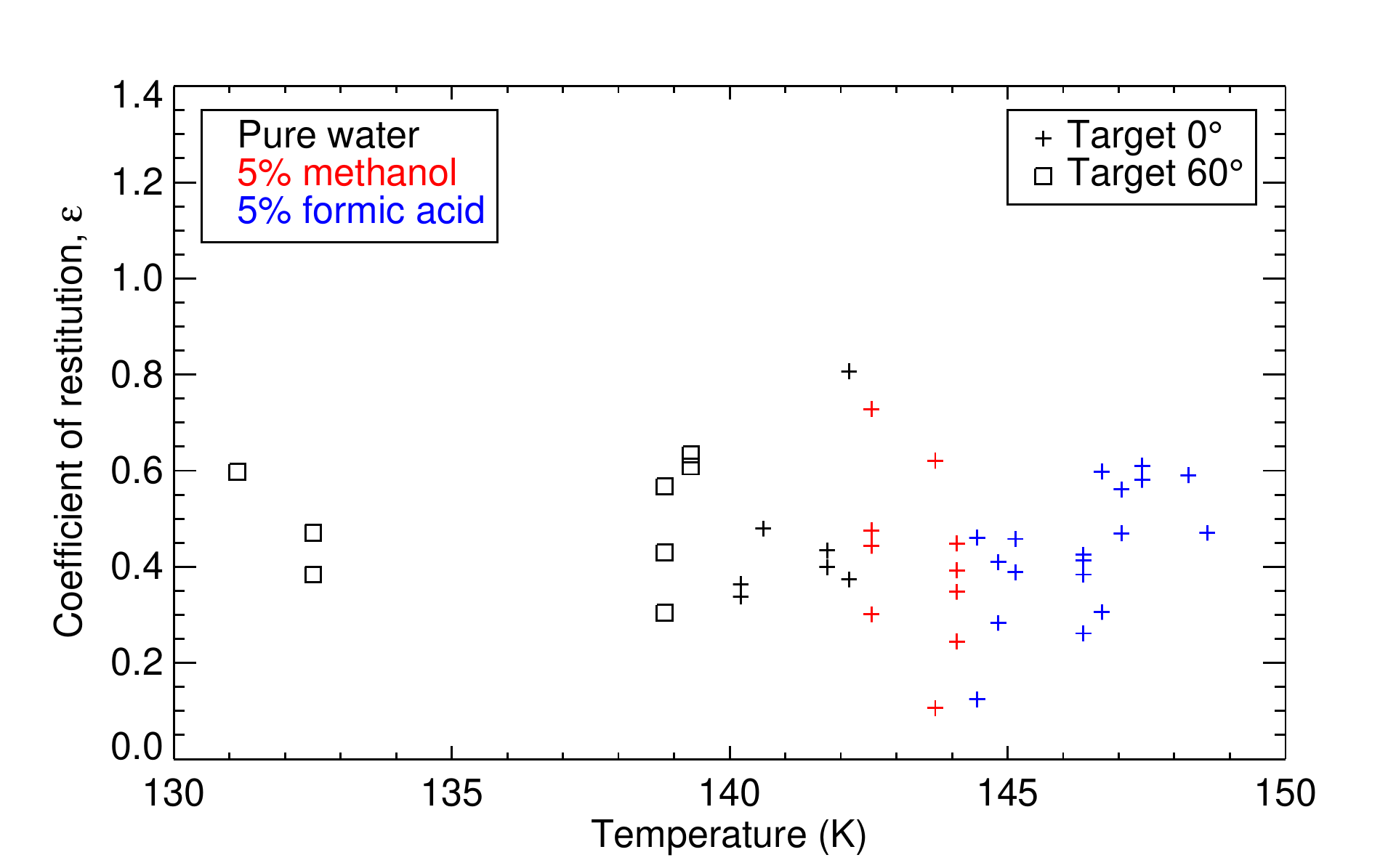}
      \caption{Coefficient of restitution as a function of temperature. There is no correlation between the two parameters. The chemical compositions are indicated by different colours: black for pure water ice, red for 5\% methanol in water, and blue for 5\% formic acid in water. The symbols indicate the orientation of the target; a plus symbol for the target at 90\ds to the direction of travel, and a square for the target at 60\ds to the direction of travel.
      }
         \label{FigCOR_T}
   \end{figure}

Due to parabolic flight restrictions, it is not possible to cool the chamber with liquid nitrogen during flight. The flow of nitrogen must be stopped prior to take off, relying on the copper in the experiment to act as a heat sink. The chamber temperature slowly increases during the flight, giving us an opportunity to study the effect of temperature on coefficient of restitution. The data for the third flight of this campaign have been excluded due to the possibility of additional warming when the chamber was opened to remove the target. The temperature ranged from 131 to 160 K, increasing at a rate of 22.5 K h$^{-1}$ for the first 20 minutes and at a rate of 7.0 K h$^{-1}$ for the remainder of the flight. Fig.~\ref{FigCOR_T} shows the coefficient of restitution as a function of temperature. The data for the second flight (target at 60\dsns) is at lower temperatures than the data for the first flight (target at 90\dsns) because the available collisions from this flight all took place earlier in the flight than those on the first flight. There is no correlation between the parameters (a correlation coefficient of -0.06) and hence we conclude that temperature has no effect on coefficient of restitution over this temperature range. This corroborates our previous work in Paper I and again we conclude that this is due to the lack of surface melting at these temperatures and pressures. In order for surface melting to take place, higher temperatures and pressures would be required (around 250 K and a few mbar, considering the phase diagram of water \citep{Ehrenfreund03}). In this temperature and pressure regime, we would expect to see a reduction in coefficient of restitution with increasing temperature owing to greater sticking forces at higher temperatures, as seen by \cite{Supulver97}.

\section{Astrophysical implications}

It appears that inclusion of methanol and formic acid at a level of 5\% has no impact on the collisional properties. However, methanol freeze out in planetary rings and protoplanetary disks will happen at lower temperatures than those at which our freezing process took place, which could lead to a truly mixed ice. In this scenario, methanol may have a greater effect on collisional properties, particularly if it is present in higher concentrations near the ice surface. However, formic acid has not been detected above the 5\% level and is likely to be evenly distributed throughout our ice sample, leading us to conclude that the inclusion of formic acid will not affect the collisional properties of the ices. Therefore, it seems unlikely that the inclusion of other species on the level of 5\% will affect the collisional outcomes of icy particles in ring systems or protoplanetary disks.

Both the pure ice spheres and the ice spheres containing 5\% methanol or formic acid yielded a broad range of coefficients of restitution. This corroborates previous results by \cite{Heisselmann10} and Paper I. Historically, the coefficients of restitution determined by \cite{Bridges84} have been most commonly employed in numerical simulations of ring system dynamics. However, our results show that since both elastic and inelastic collisions take place concurrently, it is more appropriate to use a broad range of coefficients of restitution, evenly spread between 0.08 and 0.81. This conclusion is valid for both pure ice spheres and ice spheres containing 5\% methanol or formic acid.

Although surface melting as a sticking mechanism is not possible at these temperatures and pressures for crystalline ice, a similar surface restructuring mechanism is likely to be possible for amorphous solid water, which has been detected in star forming regions \citep{Smith11} and around young stellar objects \citep{Schegerer10}, and is speculated to be present in protoplanetary disks. Amorphous solid water undergoes a glass transition to cubic ice at 137 K \citep{Smith99} which is within the temperature range of our experiments. Collisions at this temperature would be more likely to lead to sticking as bonds could be formed between the particle surfaces as they undergo restructuring and this has already been proposed as a sticking mechanism \citep{Supulver97}.


\section{Conclusions}

The main conclusions of our work are as follows:

\begin{enumerate}

\item Sticking does not occur as a collisional outcome. Universal bouncing was observed, demonstrating that the critical velocity for the onset of bouncing of ice particles of this size (1.5 cm diameter) is less than 0.01 \msns.

\item No difference was observed between the coefficients of restitution for pure crystalline water ice, water with 5\% methanol and water with 5\% formic acid. The presence of methanol and formic acid in the ice does not affect the collisional behaviour in our experiments. We postulate that this is because all samples have surface structures that are dominated by crystalline water ice. As in previous studies (Paper I and \cite{Heisselmann10}), the surface roughness is the dominant feature of the particles, not their chemical composition.

\item A broad range of coefficients of restitution was found with no correlation between this and impact velocity, replicating the results of previous studies (Paper I and \cite{Heisselmann10}). As before, this is thought to be due to the rough, anisotropic surfaces of the ice particles.

\item At least 29\% of the particles' initial translational kinetic energy is lost in the collision. Some of this energy is converted into rotational energy but this cannot account for all of the energy loss.

\item Temperature did not affect coefficient of restitution over the range measured (131 to 160 K).

\end{enumerate}


\begin{acknowledgements}
We thank the Deutsches Zentrum f\"ur Luft- und Raumfahrt (DLR) for providing us with the parabolic flights and for financial aid through grant nos. 50WM0936 and 50WM1236. We thank the European Space Agency (ESA) for providing us with the ICAPS high-speed camera system during the parabolic flight campaign. D. Hei{\ss}elmann was supported by the Deutsche Forschungsgemeinschaft (DFG) through grant no. BL 298/11-1. We would like to thank the European Community's Seventh Framework Programme LASSIE FP7/2007-2013 (Laboratory Astrochemical Surface Science in Europe) for funding H. J. Fraser and C. R. Hill's participation in the data analysis and interpretation from this work, under grant agreement no. 238258. C. R. Hill thanks The Open University for a PhD studentship; H. J. Fraser thanks SUPA (Scottish Universities Physics Alliance) and The University of Strathclyde for supporting the original experimental work. Last, but not least, we thank all parabonauts (Katharina Bairlein, Eike Beitz, Ren\'{e} Weidling and Martin Uhrin) who participated in the parabolic-flight campaign for their help in executing the experiments and Bob Dawson for his assistance during the campaign.

\end{acknowledgements}

\bibliographystyle{aa}
\bibliography{ice2}

\end{document}